\documentclass[
  notitlepage,
  preprintnumbers,
  longbibliography,
  superscriptaddress,
  aps,
  prl,
  amsmath,
  amssymb,
]{revtex4-2}

\usepackage{graphicx}
\usepackage{dcolumn}
\usepackage{bm}
\usepackage{subcaption}
\usepackage{amsmath}  

\usepackage{float}  
\begin{document}


\title{Quantum Quench Dynamics in an Exactly Solvable Two-Dimensional Non-Fermi Liquid System}

\author{Hui Li}
 \altaffiliation{Institute for Advanced Study in Physics and School of Physics, Zhejiang University, Hangzhou 310027, China}
  \email{physicslihui@zju.edu.cn}

\author{Run-Yu Chen}
 \altaffiliation{Institute for Advanced Study in Physics and School of Physics, Zhejiang University, Hangzhou 310027, China}
  \author{Wen-Yuan Liu}
 \altaffiliation{Institute for Advanced Study in Physics and School of Physics, Zhejiang University, Hangzhou 310027, China}
   \author{Yin Zhong}
 \altaffiliation{School of Physical Science and Technology \& Key Laboratory for Magnetism and
Magnetic Materials of the MoE, Lanzhou University, Lanzhou 730000, China}
 \author{Hai-Qing Lin}
 \altaffiliation{Institute for Advanced Study in Physics and School of Physics, Zhejiang University, Hangzhou 310027, China}

\date{\today}

\begin{abstract}
Understanding the behavior of non-Fermi liquids (NFLs) is an important topic in condensed matter physics. Here we introduce an exactly solvable multi-orbital model based on iron oxypnictides and the Hatsugai-Kohmoto model, and provide exact investigations of the 2D NFLs nonequilibrium physics present in this model. Our results reveal fundamental departures from Fermi liquids and prior NFLs in the well-know SYK model: anomalous short-time scaling \(-\tau^2 \ln \tau\), \(O(\tau) \sim \tau^2\); long-time scaling  \( \tau^{-1}, \tau^{-1/2}, \ln \tau / \tau\); a strange critical behavior in the steady-state phase diagram. Our asymptotic results and dynamical critical behavior offer new insights into the orbital-related dynamical physics of 2D NFLs.
\end{abstract}

\maketitle

\textit{Introduction}--
One of the central frontiers in condensed-matter physics lies in understanding non-Fermi liquids (NFLs)—strongly correlated states exhibiting a fundamental breakdown of Landau's quasiparticle paradigm\cite{landau1959theory, lee2018review}. These states emerge in various classes of correlated materials, such as cuprate\cite {WXG2006_RMP,keimer2015quantum}, iron-based superconductors\cite{analytis2014transport,paglione2010high,stewart2011superconductivity,fernandes2022iron}, and heavy-fermion compounds\cite{stockert2011unconventional,kirchner2020colloquium}, presenting a wealth of exotic phenomena.

A significant yet poorly understood challenges in NFLs physics lies in its anomalous non-equilibrium physics, which starkly contrasts with conventional Fermi liquids\cite{Else2021,jarrell1997non,dora2011luttinger,dupays2024exact,qiu2025syk,haldar2020SYK,larzul2022SYK,bandyopadhyay2023universal,banerjee2017solvable,jang2023prethermalization,jaramillo2025thermalization,chowdhury2025first,zhang2020entanglement,cheng2025hydrodynamic}. 
To date, the Sachdev-Ye-Kitaev (SYK) model\cite{sachdev1993gapless,Kitaev2015_Holography} and its variants have provided the most extensively studied analytical framework for NFLs dynamics, exhibiting signatures such as Planckian scaling and unconventional exponential or super-exponential decays—distinct from Fermi liquid prethermal plateaus\cite{banerjee2017solvable,jang2023prethermalization,bandyopadhyay2023universal,jaramillo2025thermalization}, and achieving maximal chaos through saturation of the quantum Lyapunov exponent bound\cite{qiu2025syk,haldar2020SYK,larzul2022SYK,bandyopadhyay2023universal,banerjee2017solvable,jang2023prethermalization,jaramillo2025thermalization,chowdhury2025first,zhang2020entanglement,cheng2025hydrodynamic}.
However, the SYK model is confined to zero dimensions and features a single flavor, fundamentally limiting its capacity to describe realistic correlated materials. This presents a major gap: while realistic NFLs materials (e.g., cuprates, iron-based superconductors) are usually multi-orbital systems defined on two-dimensional lattices, rigorous analytical understanding of dynamical physics in such higher-dimensional, multi-orbital settings remains largely unexplored.

To address this gap and advance the study of the 2D and multi-orbital effects of dynamics in NFLs, we propose a new model by combining the minimal two-band model of iron oxypnictides\cite{zhang2008} with the Hatsugai-Kohmoto (HK) model, a framework describing NFLs with long-range momentum-space interactions\cite{hatsugai1992hk,hatsugai1996hk,bacsi2025HK,guerci2025hk,tenkila2025hk,zhong2022solvable,zhongyin2023,zhongyin2025review}. This exactly solvable model captures both the physics of 2D NFLs and multi-orbital effects, enables detailed analytical and numerical investigations of NFLs dynamical physics.

In this article, we investigate dynamics behaviors following a sudden quench from unentangled initial states. To characterize the orbital dynamics in 2D NFLs, we focus on two key quantities: the entanglement entropy between the two orbitals, and the orbital order—defined as the difference in particle numbers between the two orbitals. A fundamental distinction arises between momentum-local and global quantities: while the evolution of single-momentum quantities is periodic, the distribution pattern of physical quantities in the first Brillouin zone ultimately stabilizes into a butterfly-like structure during long-time evolution.

The relaxation process evolves through three characteristic temporal stages: the ultra-short time region, the immediate region, and the long-time region. At ultra-short times ($\tau \to 0^+$), exact asymptotic analyses show that the orbital order obeys $ O(\tau ) \propto \tau ^2$, while the entanglement entropy follows $S(t) \propto \tau ^2 \ln(1/\tau )$.  In the intermediate regime, all global physical quantities exhibit stable oscillations with a decay that follows neither a power law nor an exponential law. In the long-time limit ($\tau  \to \infty$), orbital order decays algebraically as $\tau ^{-1}$ or $\tau ^{-1/2}$,whereas the entanglement entropy approaches its steady value via $S(\tau )  \sim \tau ^{-1}$, $\tau ^{-1/2}$, or $\ln \tau  / \tau $. 
These exact scaling laws, distinct from both Fermi liquids and SYK models, provide a concrete example of relaxation pathways of 2D NFLs. Furthermore, the global steady-state values of both orbital order and entanglement entropy reveal a critical line along \(U=V\) in the phase diagram of the steady state—offering a significantly different landscape from equilibrium ground-state phase diagrams.
Collectively, these results offer a valuable case study and new perspectives on the non-equilibrium behavior of 2D, multi-orbital NFLs relevant to quantum materials.

\begin{figure}[!htb]
    \centering
    \includegraphics[width=0.6\textwidth]{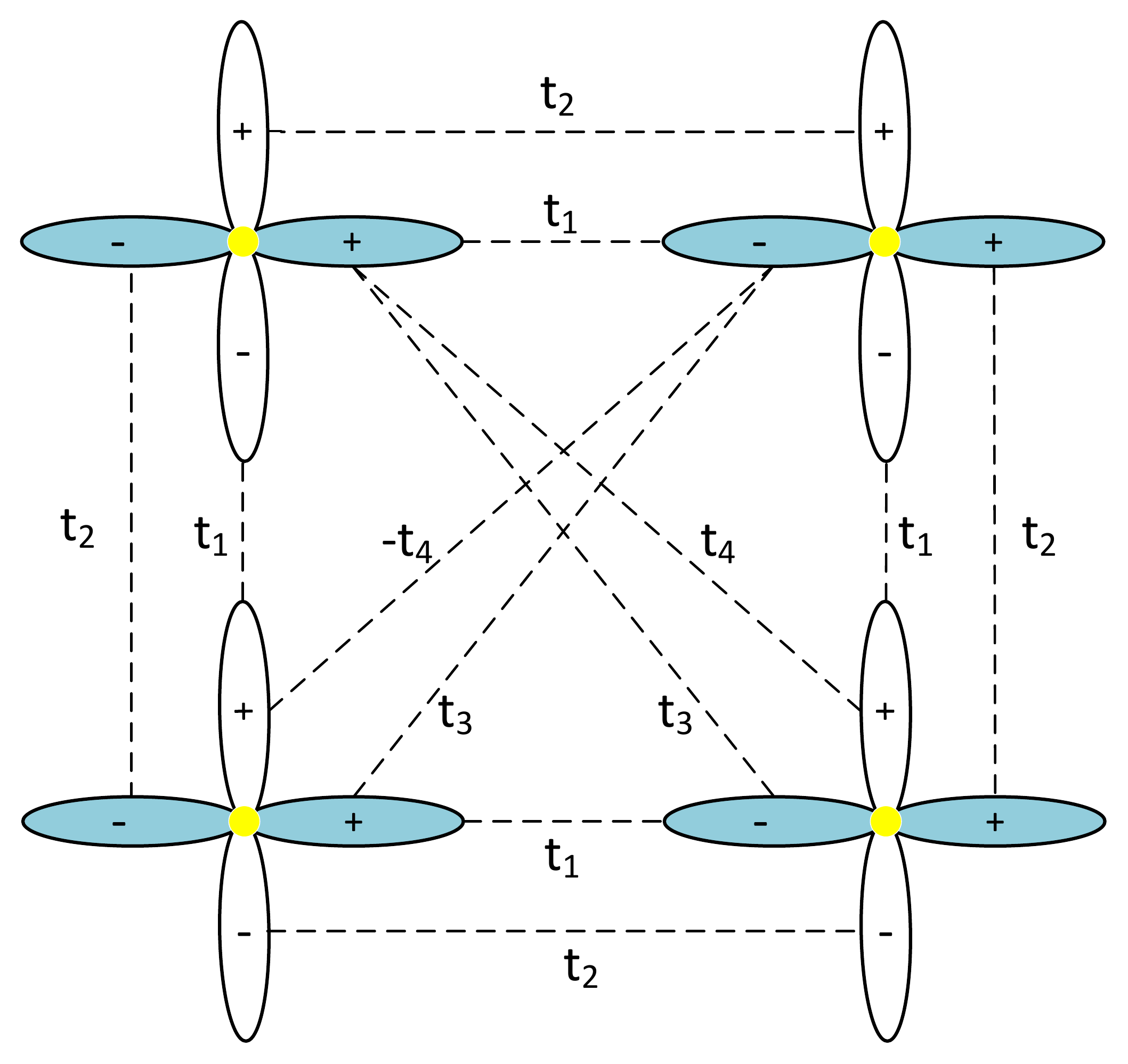}
    \caption{Schematic illustration of the hopping parameters for the two-orbital $d_{\mathrm{xz}},d_{\mathrm{yz}}$ model on a square lattice.}
    \label{fig:lattice}
 \end{figure}

\textit{Model Hamiltonian}--
In this article, we introduce a variant of the minimal two-band model of iron-based superconductors\cite{zhang2008} by including the HK interaction\cite{zhongyin2025review},
  \begin{align}
\hat H=&\sum_{\mathbf k\sigma}\sum_{\alpha\alpha'}\varepsilon_{\alpha\alpha'}(\mathbf k) \hat c_{\mathbf k\alpha\sigma}^\dagger \hat c_{\mathbf k\alpha'\sigma}\label{eq:Hamiltonian}\\
&+U\sum_{\mathbf k \alpha}  \hat n_{\mathbf k\alpha\uparrow}\hat n_{\mathbf k\alpha\downarrow}+V\sum_{\mathbf k }\hat n_{\mathbf k\mathrm{x}}\hat n_{\mathbf k\mathrm{y}}\nonumber
.
\end{align}
Here we consider a $L\times L$ square lattice with two degenerate $d_{\mathrm{xz}}$ and $d_{\mathrm{yz}}$ orbitals per site, labeled by $\alpha=\mathrm x,\mathrm y$ as shown in Fig.~\ref{fig:lattice}
The operator $\hat c^\dagger_{\mathbf k\alpha\sigma}$ creates an electron with momentum $\mathbf{k}=(k_x,k_y)$, spin $\sigma$, and orbital index $\alpha$. And $\hat n_{\mathbf k\alpha\sigma}=\hat c^\dagger_{\mathbf k\alpha\sigma}\hat c_{\mathbf k\alpha\sigma}$ is the momentum-dependent density operator with spin $\sigma$ at orbital $\alpha$. The noninteracting dispersion is 
\begin{align}
\varepsilon_{\mathrm{xx}}(\mathbf k)&=-2t_1\cos k_x-2t_2\cos k_y -4t_3\cos k_x\cos k_y,\nonumber\\
\varepsilon_{\mathrm{yy}}(\mathbf k)&=-2t_1\cos k_y-2t_2\cos k_x -4t_3\cos k_x\cos k_y,\nonumber\\
\varepsilon_{\mathrm{xy}}(\mathbf k)&=-4t_4\sin k_x \sin k_y,
\end{align}
where $t_1$ quantifies the strength of the hoping from $d_{\mathrm{xz}}$ to $d_{\mathrm{xz}}$ along the $x$ direction,  and also $d_{\mathrm{yz}}$ to $d_{\mathrm{yz}}$ hopping along the $y$ direction. $t_2$ describes the hopping strength from $d_{\mathrm{xz}}$ ($d_{\mathrm{yz}}$) to $d_{\mathrm{xz}}$ ($d_{\mathrm{yz}}$) along the $y$ ($x$) direction. Additionally, we include the second-neighbor hopping $t_4$ between different orbitals and $t_3$ between the same orbitals. $U$ is the intra-orbital interaction, $V$ the inter-orbital ($d_{\mathrm{xz}}$-$d_{\mathrm{yz}}$) interaction, and $\hat n_{\mathbf{k}\alpha} = \hat n_{\mathbf{k}\alpha\uparrow} + \hat n_{\mathbf{k}\alpha\downarrow}$. Unless otherwise specified, we take the hopping parameters in our calculations as follows: $t_1=-1,t_2=1.3,t_3=t_4=0.85$. 

By introducing degenerate $d_{\mathrm{yz}}/d_{\mathrm{xz}}$ orbitals with anisotropic hopping, we construct a minimal model supporting orbital-resolved non-equilibrium dynamics while remaining exactly solvable. Crucially, the inter-orbital interaction (V) combined with orbital-dependent hopping enables momentum-dependent charge transfer between orbitals, breaking the local frozen dynamics characteristic of single-band models. This model thus provides an exactly solvable platform for studying 2D interaction-driven quantum dynamics of NFLs beyond mean-field approximations.

The key feature of the Hamiltonian Eq.~(\ref{eq:Hamiltonian}) is the momentum-space locality $\hat H = \sum_{\mathbf{k}} \hat H_{\mathbf{k}}$. This "local" Hamiltonian can be diagonalized in the Fock space $\left| \{n \} \right> =\left| n_{x\uparrow}, n_{x\downarrow}, n_{y\uparrow}, n_{y\downarrow} \right>_k$, which can be constructed as
\begin{equation}
   \left| \{n\} \right>_{\mathbf k}=(\hat c_{\mathbf{k}\mathrm{x}\uparrow})^{ n_{\mathrm{x}\uparrow}}(\hat c_{\mathbf{k}\mathrm{x}\downarrow})^{ n_{\mathrm{x}\downarrow}}(\hat c_{\mathbf{k}\mathrm{y}\uparrow})^{ n_{\mathrm{y}\uparrow}}(\hat c_{\mathbf{k}\mathrm{y}\downarrow})^{ n_{\mathrm{y}\downarrow}}\left| 0 \right>,
\end{equation}
where each $n$ takes the value of 0 or 1. At each $\mathbf{k}$, $\hat H_{\mathbf{k}}$ forms a $16\times16$ block matrix separable into 9 subspaces $\left| n_{\uparrow}, n_{\downarrow} \right>$ due to spin conservation, with $n_\uparrow = n_{x\uparrow} + n_{y\uparrow}$ and $n_\downarrow = n_{x\downarrow} + n_{y\downarrow}$ ($n_{\uparrow,\downarrow} = 0,1,2$). Numerical diagonalization of $\hat H_{\mathbf{k}}$ at each momentum yields eigenvalues $E_n(\mathbf{k})$ and corresponding eigenstates $\left| \phi_n(\mathbf{k}) \right>$. 
Since the Hamiltonian is decoupled in momentum space, the full eigenstate $\left| \Phi \right>$ can be constructed as a direct product of momentum-resolved states:
\begin{equation}
\left| \Phi \right> = \bigotimes_{\mathbf{k}} \left| \phi_\mathbf{k} \right>,
\end{equation}
where $\left| \phi_\mathbf{k} \right>$ denotes the eigenstate at momentum $\mathbf{k}$.

\textit{Dynamics for physical quantites}--
Due to the Schrödinger equation, for a pure state initialized in $\left| \Psi \right>=\prod_{\mathbf k} \bigotimes \left| \psi_{\mathbf k} \right>$, the state evolves as
\begin{align}
  \left| \Psi(\tau ) \right>&=e^{-\mathrm{i}\hat H \tau }\left| \psi \right>=\prod_k\bigotimes e^{-\mathrm{i}\hat H_k \tau }\left| \psi_{\mathbf{k}} \right>.\label{eq:quench_operator}
\end{align}
In this article, we investigate the orbital order operator $\hat O $ and the inter-orbital entanglement entropy between $d_{\mathrm{xz}}$ and $d_{\mathrm{yz}}$ orbitals. The orbital order $\hat O=\oplus_k\hat O_{\mathbf k}$ can be measured as
\begin{equation}
O(\tau ) = \left\langle \hat{O}(\tau ) \right\rangle 
= \sum_{\mathbf{k}} \bigl\langle \psi_{\mathbf{k}}(\tau ) \mid \hat{O}_{\mathbf{k}} \mid \psi_{\mathbf{k}}(\tau ) \bigr\rangle,\label{eq:evolution}
\end{equation}
where $\hat O_{\mathbf k}=\hat n_{\mathbf{k}\mathrm{x}}-\hat n_{\mathbf{k}\mathrm{y}}$ is the momentum-dependent orbital order.
The momentum-resolved entanglement entropy $S_{\mathbf{k}}(\tau)$ is computed from the reduced density matrix $\hat \rho_{\mathbf{k}\mathrm{x}}(\tau) = \mathrm{tr}_{\mathrm{y}} \left( \left| \psi_{\mathbf{k}}(\tau) \right\rangle \left\langle \psi_{\mathbf{k}}(\tau) \right| \right)$ as:
\begin{equation}
S_{\mathbf{k}}(\tau) = -\mathrm{tr} \left( \hat\rho_{\mathbf{k}\mathrm{x}}(\tau) \ln \hat\rho_{\mathbf{k}\mathrm{x}}(\tau) \right).
\end{equation}
The global physical quantities are then obtained by momentum-space averaging:
\begin{align}
S(\tau) = \frac{1}{L^2} \sum_{\mathbf{k}} S_{\mathbf{k}}(\tau) , \quad
O(\tau) = \frac{1}{L^2} \sum_{\mathbf{k}} O_{\mathbf{k}}(\tau) .
\end{align}

\textit{Numerical Dynamical calculation}--We initiate the dynamics from a pure state within the $\{ n_\uparrow = 1, n_\downarrow = 1\}$ subspace of dimension 4, specifically prepared with double occupation in the $d_{\mathrm{xz}}$-orbital and vacuum in the $d_{\mathrm{yz}}$-orbital for each $\mathbf k$:
\begin{equation}
|\psi_{\mathbf k,\mathrm{initial}}\rangle = |n_{x\uparrow}{=}1, n_{x\downarrow}{=}1, n_{y\uparrow}{=}0, n_{y\downarrow}{=}0\rangle.
\end{equation}
This configuration maximizes dynamical richness while preserving spin and charge conservation, essential for probing orbital-resolved many-body effects. Then we apply a sudden quench described by the Eq.~(\ref{eq:Hamiltonian},\ref{eq:quench_operator}) to study the dynamical process. Here we study the following three-types of characters in the dynamical physics: (i) the dynamical evolution for momentum-dependent quantities $O_{\mathbf k}(\tau),S_{\mathbf k}(\tau)$; (ii) the dynamical evolution for global quantities $O(\tau),S(\tau)$; (iii) the global quantities at the steady states for long-time limit $\bar O,\bar S$.

\begin{figure}[htbp]
\centering
\includegraphics[width=0.6\columnwidth]{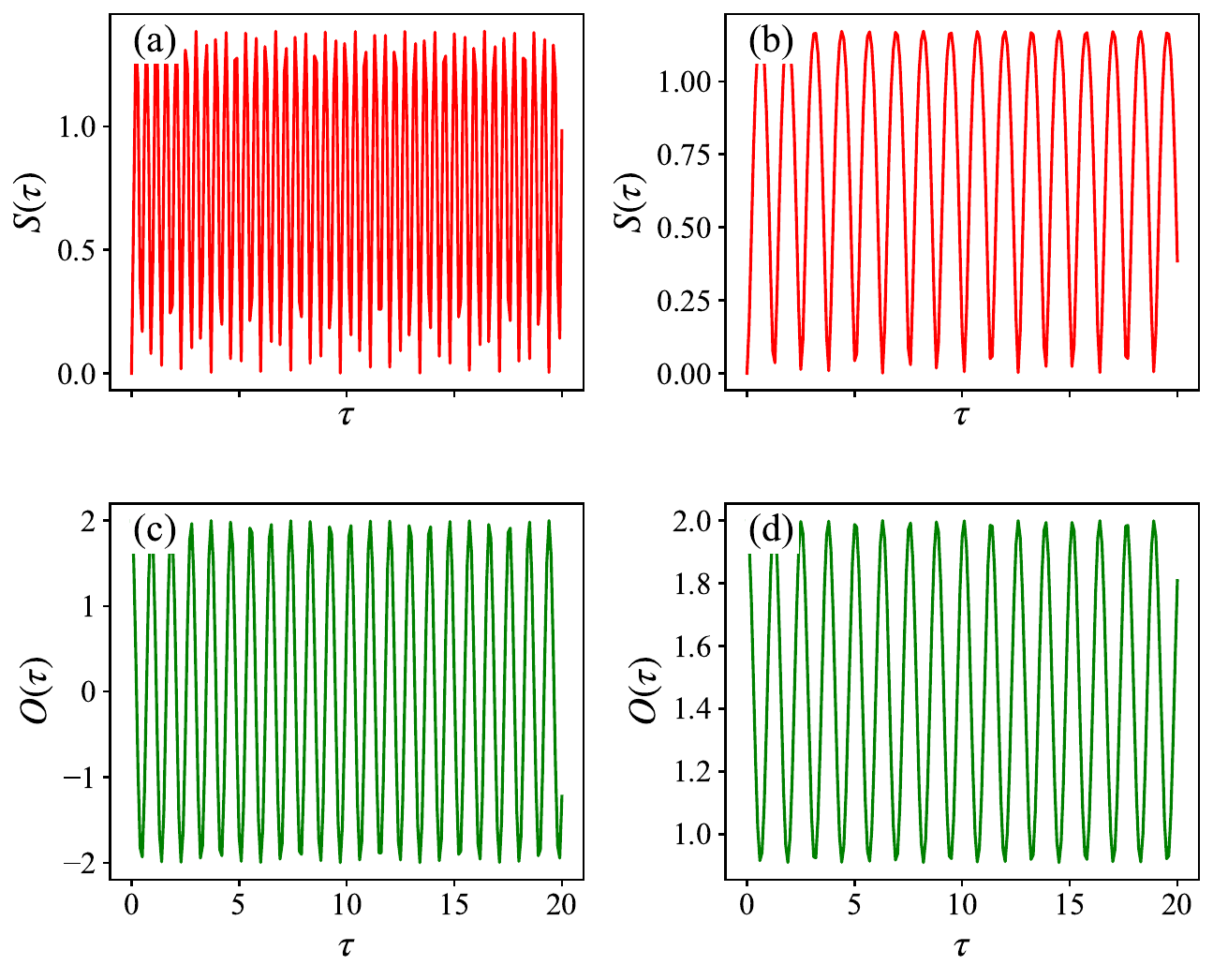}
\caption{Persistent oscillations in $S_{\mathbf k}(\tau )$ and $O_{\mathbf k}(\tau )$ at two representative momenta: (a)(c) $\mathbf{k} = (\pi/2,\pi/2)$; (b)(d) $\mathbf{k} = (\pi/2,\pi/8)$}
\label{fig:single_momentum}
\end{figure}

\textit{The dynamics for momentum-dependent quantities}--Post-quench dynamics reveals a fundamental dichotomy between local quantum periodicity and global thermalization, as demonstrated through momentum-resolved analysis. Fig.~\ref{fig:single_momentum} shows periodic oscillations in both entanglement entropy \( S_{\mathbf{k}}(\tau) \) and orbital order \( O_{\mathbf{k}}(\tau) \) at isolated momentum points \( \mathbf{k} = (\pi/4,\pi/4) \) and \( \mathbf{k} = (\pi/8,\pi/4) \), where oscillation frequencies differ for each momentum. This robust momentum-dependent periodicity originates from the finite effective 4-dimensional Hilbert space per \(\mathbf{k}\)-point, as governed by Eq.~(\ref{eq:quench_operator}).

\begin{figure*}[htb]
    \centering
    \includegraphics[width=0.9\textwidth]{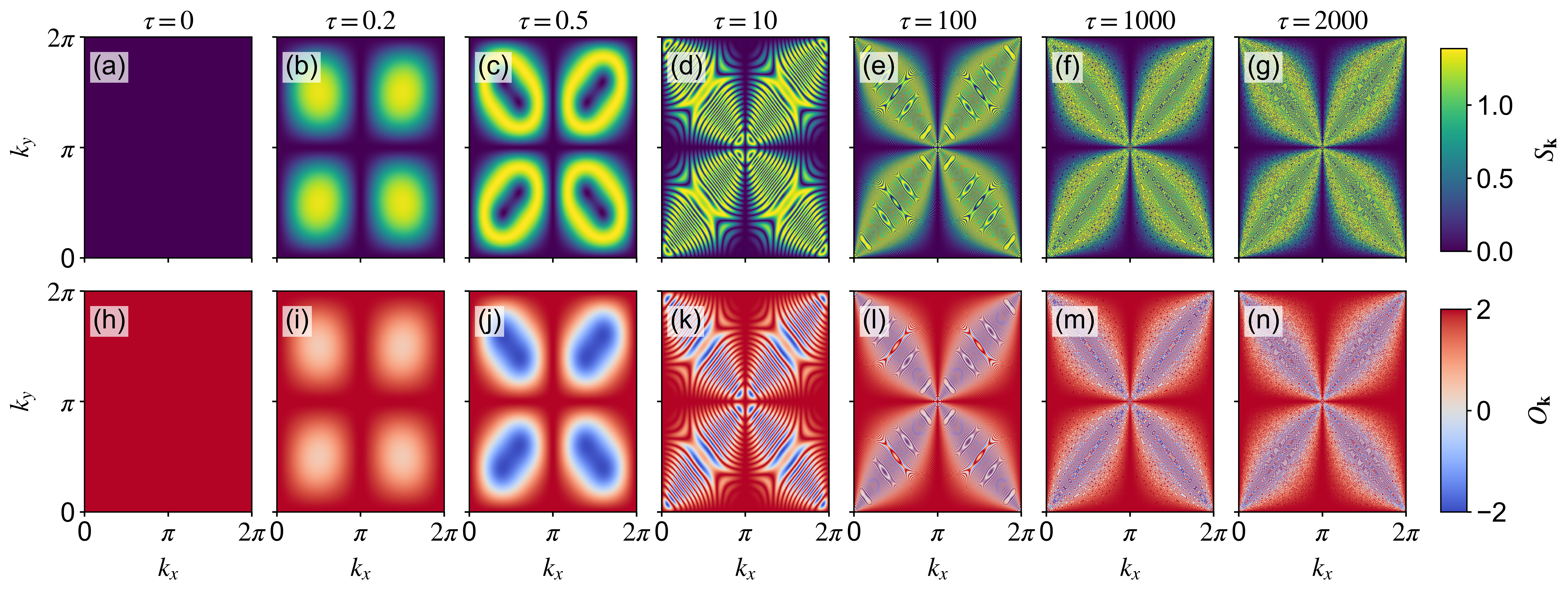}
    \caption{Momentum-space dynamical patterns for entanglement entropy (a–g) and orbital order (h–n) at selected time points. (a)(h) at $\tau =0$;(b)(i) at $\tau =0.2$, (c)(j) at $\tau =0.5$, (d)(k) at $\tau =10$ shows the relaxation process; (a)(h) at $\tau =0$ shows the un-quenched situation;(c)(l) at $\tau =100$, (f)(m) at $\tau =1000$, (g)(n) at $\tau =2000$ display the stable butterfly-like pattern.}
    \label{fig:pattern}
\end{figure*}

Different from the momentum-dependent periodicity, the distribution pattern in the momentum space displays an irreversible formation, as shown sequentially in Fig.~\ref{fig:pattern}. Starting from a homogeneous, featureless distribution at $\tau =0$, the system rapidly forms a symmetric ring structure post-quench, which subsequently ($\tau  > 10$) decays into a non-fractal butterfly-like pattern. This pattern exhibits a distinctive $\sin k_x \sin k_y$ symmetry, inherited directly from the orbital hybridization energy $\varepsilon_{\mathrm{xy}} = -t_4 \sin k_x \sin k_y$. The pattern's stability stems from momentum-dependent variations in both oscillation amplitudes and frequencies: amplitudes peak at $\mathbf{k}=(\pi/2,\pi/2)$ where $|\sin k_x \sin k_y|$ maximizes but small at edges like $\mathbf{k}=(\pi/2,\pi/8)$ (Fig.~\ref{fig:single_momentum}), while frequencies differ significantly across $\mathbf{k}$-points. This amplitude-frequency decoherence drives entanglement entropy and orbital order to dynamical equilibrium via destructive interference, making the geometric pattern a direct signature of dynamical orbital effect.

\begin{figure}[htb]
    \centering
    \includegraphics[width=0.7\textwidth]{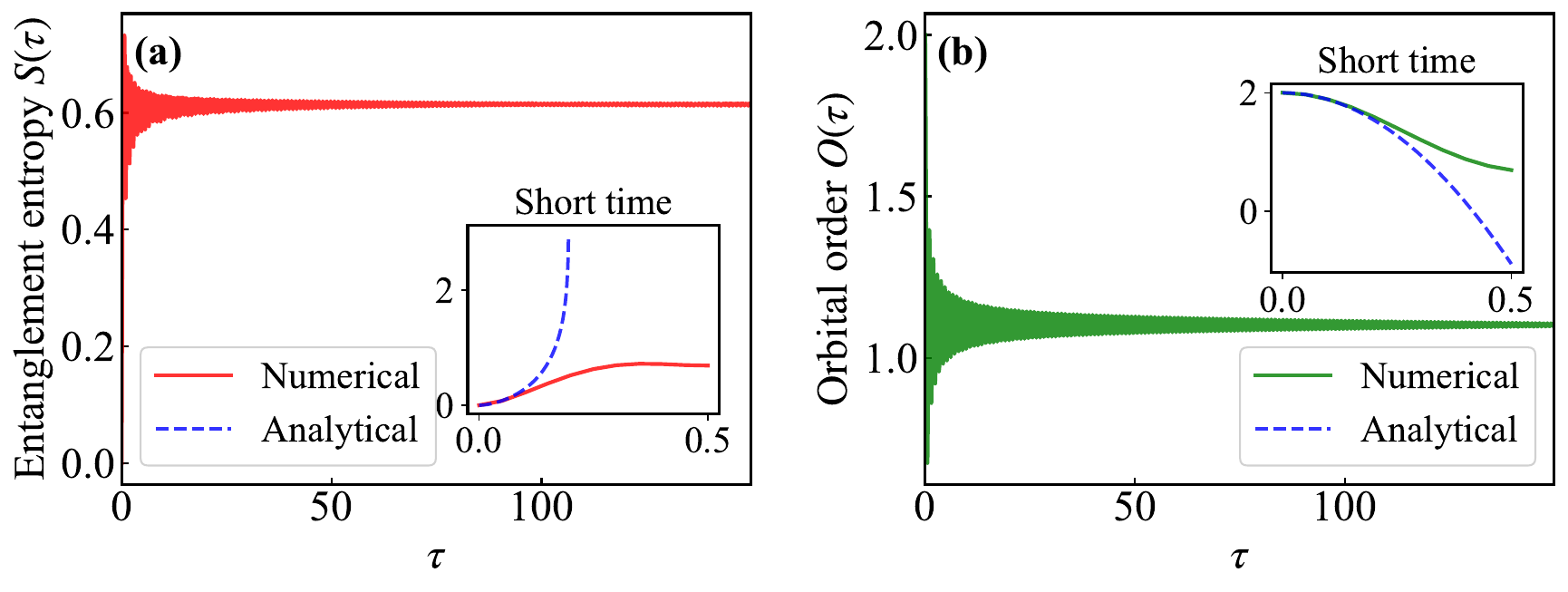}
    \caption{The relaxation process of (a) entanglement entropy $S(\tau )$, (b) orbital order $O(\tau )$ for $U=V$ at $1216\times1216$ lattice. Insert: the short-time evolution of each quantity, the dashed blue line labels the analytic predictions following Eq.~\ref{eq:short-time}.}
    \label{fig:relax}
\end{figure}

\textit{The dynamics for global quantities}--We investigate the relaxation asymptotic scaling of physical quantities. As shown in Fig.~\ref{fig:relax}, both the global entropy $S(\tau)$ and the orbital order $O(\tau)$ exhibit distinct three-stage relaxation dynamics: (1) rapid initial evolution, (2) persistent oscillations in the intermediate stage characterized by complex non-monotonic decay (neither power-law nor exponential), and (3) asymptotic decay governed by a well-defined scaling law towards their steady-state values.

In the short-time limit ($\tau  \to 0^+$), evolution operator can be expanding as
\begin{equation}
    e^{-\mathrm{i}\hat H \tau }\approx \hat I-\mathrm{i}\hat H \tau -\frac{1}{2}\hat H^2 \tau ^2,
\end{equation}
reveals that both $O(\tau )$ and $S(\tau )$ exhibit parameter-independent universal scaling laws:
\begin{align}
O(\tau ) \approx 2 - 16t_4^2 \tau ^2, \quad S(\tau ) \approx -16t_4^2 \tau ^2 \ln \tau .\label{eq:short-time}
\end{align}
This universal behavior, confirmed numerically in Fig.~\ref{fig:relax}, indicates the inter-orbital hopping $t_4$ dominating the early-time dynamics. 

In contrast, the long-time asymptotic behavior ($\tau  \to \infty$) displays a rich variety of scaling laws, highly sensitive to orbital symmetry ($t_1,t_2$) and interaction difference $\delta U$, as summarized in Table.~\ref{tab:scaling}. The key findings are:

\begin{itemize}
    \item Symmetric case ($t_1 = t_2$): 
    \begin{itemize}
        \item For $\delta U = 0$, When $\delta U = 0$, orbital order can be solved analytically as $O(\tau ) = 2[J_0(4t_4\tau )]^2$, which decays as $O(\tau ) \sim \tau ^{-1}$, and entanglement entropy as $S(\tau ) \sim (\ln \tau )/\tau $. 
        \item For $\delta U \neq 0$, both quantities follow $\tau ^{-1/2}$ scaling, arising from one-dimensional stationary manifolds in momentum space.
    \end{itemize}
    
    \item Asymmetric case ($t_1 \neq t_2$):
    \begin{itemize}
        \item For $\delta U = 0$, the standard $\tau ^{-1}$ decay emerges, originating from isolated stationary points.
        \item For $\delta U \neq 0$, the decay follows $\tau ^{-1}$ or slower as $\tau ^{-\gamma}$, depending on the energy-gap criticality at stationary points. If the stationary points are degenerate, the exponent $\gamma = 1/b + 1/c$ is determined by the energy-gap expansion $\Delta E \sim k_x^b + k_y^c$ around the critical momentum. Crucially, $\gamma < 1$ leads to anomalously slow decay.
    \end{itemize}
\end{itemize}

These diverse long-time scalings are fundamentally governed by stationary-phase contributions at critical momenta where $\nabla_{\mathbf{k}}\Delta E(\mathbf{k}) = 0$. The emergence of slow decays ($\tau^{-\gamma}$ with $\gamma < 1$) occurs when $\Delta E(\mathbf{k})$ vanishes with high power-law exponents at degenerate stationary points. This rich tapestry of dynamical scaling behaviors stands in sharp contrast to the simple, universal relaxation expected in Fermi liquids. And we can find the very different dynamical scaling behaviors in 2D NFLs from traditional SYK relaxation structure.

\begin{table*}[t]
\centering
\caption{Asymptotic behaviors of orbital order $O(\tau )$ and entanglement entropy $S(\tau )$ in the quench dynamics}
\label{tab:scaling} 
\begin{tabular}{|c|c|c|c|c|}
\hline
Parameter regime & \multicolumn{2}{c|}{orbital order $O(\tau )$} & \multicolumn{2}{c|}{Entanglement entropy $S(\tau )$} \\
\cline{2-5}
 & Short-time ($\tau  \to 0^+$) & Long-time ($\tau  \to \infty$) & Short-time ($\tau  \to 0^+$) & Long-time ($\tau  \to \infty$) \\
\hline
$t_1 = t_2$, $\delta U = 0$ & $\tau ^2$ & $\tau ^{-1}$ & $-\tau ^2 \ln \tau $ & $(\ln \tau )/\tau $ \\
$t_1 = t_2$, $\delta U \neq 0$ & $\tau ^2$ & $\tau ^{-1/2}$ & $-\tau ^2 \ln \tau $ & $\tau ^{-1/2}$ \\
$t_1 \neq t_2$, $\delta U = 0$ & $\tau ^2$ & $\tau ^{-1}$ & $-\tau ^2 \ln \tau $ & $\tau ^{-1}$ \\
$t_1 \neq t_2$, $\delta U \neq 0$ & $\tau ^2$ & $\tau ^{-\gamma}$ or slower & $-\tau ^2 \ln \tau $ & $\tau ^{-\gamma}$ or slower \\
\hline
\end{tabular}
\end{table*}

\begin{figure}[h!]
    \centering
    \includegraphics[width=0.7\textwidth]{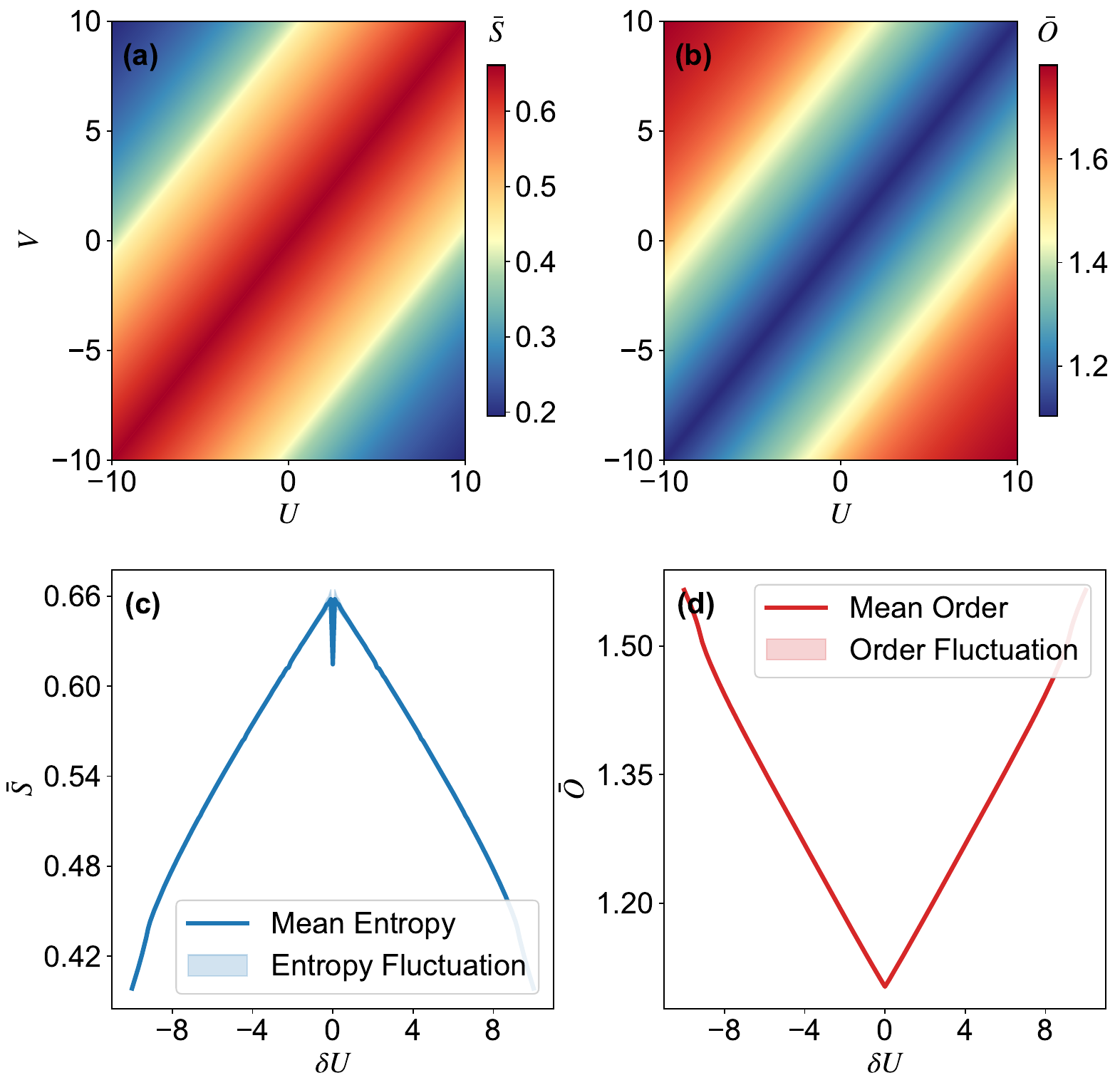}
\caption{Steady-state phase diagrams and criticality: (a)(b) Global entanglement entropy $\bar{S}$ and orbital order $\bar{O}$ in $U$-$V$ plane. (c,d) Cuts across critical line show sharp variations of $\bar{S}$ and $\bar{O}$ near $U=V$ calculated in $4096\times4096$ lattice.}
\label{fig:critical}
\end{figure}

\textit{The critical phenomenon in the steady states}--
As shown in Fig.~\ref{fig:critical}(a)(b), quench steady states exhibit distinct critical behavior in global quantities. The time-averaged entanglement entropy $\bar{S} = \lim_{T\to\infty} \frac{1}{T} \int_0^T d\tau   S(\tau )$ and orbital order $\bar{O} = \lim_{T\to\infty} \frac{1}{T} \int_0^T d\tau   O(\tau )$ converge to stationary values with fluctuations vanishing algebraically with system size. Crucially, their $U$-$V$ phase diagrams reveal a critical line at $U=V$ that is analytically solvable—a feature distinct from conventional equilibrium criticality. 

This criticality arises through the interplay of two distinct mechanisms: 
(1) The block-diagonal Hamiltonian decomposition $\hat H_{4\times4}(U,V) = \hat{\mathcal{H}}_{4\times4}(U-V) + V \hat{I}_{4\times4}$ 
(2) The emergent reflection symmetry for global observables:
\begin{equation}
O(\delta U) = O(-\delta U), \quad S(\delta U) = S(-\delta U) 
\label{eq:symmetry}
\end{equation}
where $\delta U=U-V$. Here the symmetry in Eq.~\eqref{eq:symmetry} derives from the momentum-space invariance under $\mathbf{k} \to \mathbf{k}+(\pi,\pi)$ combined with the Hamiltonian's functional dependence on $\delta U$. 

The synergy between these elements manifests as a symmetry axis at $U=V$ where observables change abruptly shown in Fig.~\ref{fig:critical}(c)(d). This atypical criticality demonstrates how symmetry constraints govern non-equilibrium critical behavior, different from the ground state phase diagram shown in the Supplementary material(SM).

\textit{Conclusion}--
In summary, we propose an exactly solvable multiorbital model, based on the model of iron oxypnictides and the HK model, to investigate the quench dynamics of 2D NFLs. Our findings reveal a fundamental distinction between local periodicity and global relaxation. While single-momentum quantities exhibit persistent periodic oscillations, the global distribution in momentum space evolves irreversibly toward a stable butterfly pattern, which present a clear signature of the mutiorbital dynamical effects.

The most distinctive nonequilibrium physics in this article lies in the temporal hierarchy of the relaxation dynamics, distinct from prior dynamical paradigms: an initial quadratic growth ($O(\tau )\sim \tau ^2$, $S(\tau )\sim-\tau ^2\ln \tau $) dominated by inter-orbital hybridization, and long-time asymptotics governed by the saddle-point of the energy gap structure, yielding interaction-tuned algebraic thermalization—from rapid $\tau^{-1}$ to anomalous $\tau^{-1/2}$ and $\ln \tau/\tau$ scaling. Compared to the super-exponential or exponential thermalization in the SYK model, this slower relaxation provides a new perspective on the dynamical mechanism of NFLs.

Furthermore, we uncover robust dynamical criticality along $U=V$ in the steady-state phase diagram, where the time-averaged orbital order $\bar O$ and entanglement entropy $\bar S$ exhibit a sharp, symmetry-driven critical line, distinct from traditional equilibrium phase structures.

As a minimal, exactly solvable theory of 2D NFLs dynamics, it establishes a benchmark for different non-equilibrium approaches, and it first provide the exact understanding in 2D NFLs dynamical orbital-related physics. Besides, the dynamical mechanism in this article can contribute to understanding the dynamics of iron-based superconductors and future experiments on 2D long-range systems.

\begin{acknowledgments}
The authors are very grateful to Sheng Yang, Lei Ying for valuable discussions.
This work is supported by MOST 2022YFA1402701. This work is also supported by the Supercomputing Center of
Lanzhou University, which provided essential computational
resources. We also acknowledge support from the National
Natural Science Foundation of China (Grant No. 12247101),
the Fundamental Research Funds for the Central Universities (Grant No. lzujbky-2024-jdzx06), the Natural Science
Foundation of Gansu Province (Grant Nos. 22JR5RA389 and
25JRRA799), and the ‘111 Center’ under Grant No. B20063.
\end{acknowledgments}
\bibliography{apssamp}

\newpage
\appendix
\section*{Supplementary Material}
\section{The phase diagram of the ground states}
\begin{figure}[H]  
    \centering
    \includegraphics[width=0.9\textwidth]{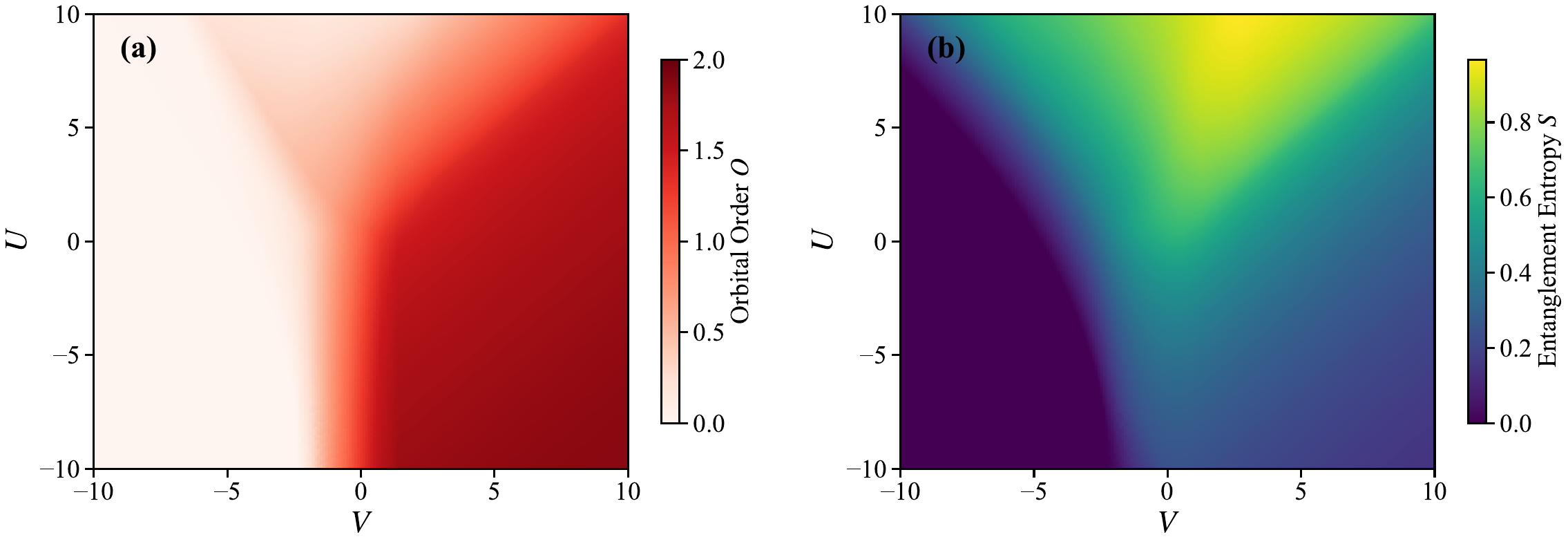}  
    \caption{Phase diagram at ground state with $n=2$ for (a) orbital order, (b) entanglement entropy}
    \label{fig:phase_diagram}  
\end{figure}

In Fig.~\ref{fig:phase_diagram}, we fix $n=2$ and establish a ground-state phase diagram for different $U$ and $V$. It shows that, unlike the symmetry structure in the dynamical phase diagram, the orbital order is large for the large U-V region, and the hybrid term $V$ plays a more important role in the phase transition. The entanglement entropy $S$ increases with larger inter-orbital interaction $V$. Conversely, $S$ may decrease as the intra-orbital interaction $U$ increases. This reveals a strikingly different phase structure between the equilibrium and non-equilibrium states.

\section{General Entanglement Entropy and Orbital Order Dynamics}

We consider the quench evolution of the HK model in the 4-dimensional subspace of $n_{k\uparrow}=1,n_{k\downarrow}=1$, with the initial state chosen as $n_{x\uparrow}=n_{x\downarrow}=1,n_y=0$. For the Hilbert subspace of $n_{k\uparrow}=1,n_{k\downarrow}=1$, the basis vectors are
\begin{equation}
\left| \uparrow_x, \downarrow_y \right>,\left| \downarrow_x,\uparrow_y \right>,\left| \uparrow\downarrow_x,0_y \right>,\left| 0_x,\uparrow\downarrow_y \right>,
\end{equation}
which we denote as $\left| n \right>$. The corresponding Hamiltonian in this subspace $h_{n_{\uparrow}=1,n_{\downarrow}=1}(k)$ can be written as
\begin{equation}
\begin{bmatrix}
\varepsilon_{xx}(k)+\varepsilon_{yy}(k)+V & 0 & \varepsilon_{yx}(k) & \varepsilon_{xy}(k)\\
0 & \varepsilon_{xx}(k)+\varepsilon_{yy}(k)+V & \varepsilon_{yx}(k) & \varepsilon_{xy}(k)\\
\varepsilon_{xy}(k) & \varepsilon_{xy}(k) & 2\varepsilon_{xx}(k)+U & 0\\
\varepsilon_{yx}(k) & \varepsilon_{xy}(k) & 0 &2\varepsilon_{yy}(k)+U\\
\end{bmatrix},
\end{equation}
where
\begin{align}
\varepsilon_{\mathrm{xx}}(\mathbf k)&=-2t_1\cos k_x-2t_2\cos k_y -4t_3\cos k_x\cos k_y,\nonumber\\ 
\varepsilon_{\mathrm{yy}}(\mathbf k)&=-2t_1\cos k_y-2t_2\cos k_x -4t_3\cos k_x\cos k_y,\nonumber\\ 
\varepsilon_{\mathrm{xy}}(\mathbf k)&=\varepsilon_{yx}=-4t_4\sin k_x \sin k_y.
\end{align}
Due to the above Hamiltonian, it can be equivalently written as
\begin{equation}
h(\mathbf k)=\begin{bmatrix}
\varepsilon_{xx}(k)+\varepsilon_{yy}(k) & 0 & \varepsilon_{yx}(k) & \varepsilon_{xy}(k)\\
0 & \varepsilon_{xx}(k)+\varepsilon_{yy}(k) & \varepsilon_{yx}(k) & \varepsilon_{xy}(k)\\
\varepsilon_{xy}(k) & \varepsilon_{xy}(k) & 2\varepsilon_{xx}(k)+\delta U & 0\\
\varepsilon_{yx}(k) & \varepsilon_{xy}(k) & 0 &2\varepsilon_{yy}(k)+\delta U\\
\end{bmatrix}+VI_{4\times 4},
\end{equation}
where $\delta U=U-V$. Since $VI_{4\times 4}$ is a constant term, it does not affect the eigenstates and energy level differences. Therefore, we will only discuss the influence of $\delta U$ below.
For the dynamical evolution with the initial state $\left| \psi_{\mathrm{initial}} \right>=\left| \left| \uparrow\downarrow_x,0_y \right> \right>=(0,0,1,0)^T$, the wave function at time $\tau$ is
\begin{equation}
\left| \psi(\tau) \right>=e^{-i\hat H \tau}\left| \psi_{\mathrm{initial}} \right>=\sum_i e^{-\mathrm{i}E_i \tau}v_{i3}\left| \psi_i \right>,
\end{equation}
where $v_{i3}=\left< \psi_i| \psi_{\mathrm{initial}} \right>$ is the component of the $i$-th eigenstate of the Hamiltonian in the third basis. The coefficients of the state vector in the original basis at time $\tau$ are
\begin{equation}
c_n(\tau)=\left< n | \psi(\tau) \right>=\sum_i e^{-\mathrm{i}E_i \tau}v_{i3}\left< n | \psi_i \right>=\sum_i e^{-\mathrm{i}E_i \tau}v_{i3}v_{in}.
\end{equation}
We define orbital order as the difference in the number of particles in $d_{\mathrm{xz}}$ and $d_{\mathrm{yz}}$. For a specific momentum in the subspace of $n_{k\uparrow}=1,n_{k\downarrow}=1$, the orbital order takes the form
\begin{equation}
\hat O=
\begin{bmatrix}
0&0&0&0\\
0&0&0&0\\
0&0&2&0\\
0&0&0&-2
\end{bmatrix}.
\end{equation}
In general, if the time-dependent wave function is of the form $\left| \psi(\tau) \right>=\sum_n c_n\left| n \right>$, the orbital order can be written as
\begin{equation}
O(\mathbf k,\tau)=2(|c_3(\tau)|^2-|c_4(\tau)|^2).
\end{equation}
For an infinite lattice system, the average orbital order in the first Brillouin zone is defined as
\begin{equation}
O(\tau)=\frac{1}{(2\pi)^2}\iint_{-\pi}^\pi O(\mathbf k,\tau)dk_x dk_y.
\end{equation}
We also consider the entanglement entropy between the two orbits $d_{\mathrm{xz}}$ and $d_{\mathrm{yz}}$. For the wave function in the subspace, generally, it can be written as
\begin{equation}
\left| \psi \right>_{\mathbf{k}}=c_1\left| \uparrow_x, \downarrow_y \right>_{\mathbf{k}}+c_2\left| \downarrow_x,\uparrow_y \right>_{\mathbf{k}}+c_3\left| \uparrow\downarrow_x,0_y \right>_{\mathbf{k}}+c_4\left| 0_x,\uparrow\downarrow_y \right>_{\mathbf{k}}\equiv\sum_nc_n\left| \varphi_n \right>_{\mathbf{k}}.
\end{equation}
Accordingly, the density matrix can be written as
\begin{equation}
\rho=\left| \psi \right>\left< \psi \right|=\sum_{mn}c_m^*c_n\left| \varphi_n\right>\left< \varphi_m \right|.
\end{equation}
The reduced density matrix has a special structure in this case:
\begin{equation}
\rho_x=\mathrm{tr}_y\rho=|c_1|^2\left| \uparrow_x\right>\left< \uparrow_x \right|+|c_2|^2\left| \downarrow_x\right>\left< \downarrow_x \right|+|c_3|^2\left| \uparrow\downarrow_x\right>\left< \uparrow\downarrow_x \right|+|c_4|^2\left| 0_x\right>\left< 0_x \right|.
\end{equation}
Thus, the entanglement entropy is
\begin{equation}
S=-\sum_n|c_n|^2\log|c_n|^2.
\end{equation}
Similarly, the average entanglement entropy within the first Brillouin zone is
\begin{equation}
S(\tau)=\frac{1}{4\pi^2}\int\int_{-\pi}^\pi S(\mathbf k,\tau)dk_x dk_y.
\end{equation}

\section{Short-Time Evolution and Asymptotic Behavior}

In the short-time evolution, the evolution operator can be approximated as
\begin{equation}
e^{-i\hat H \tau}\approx I-i\hat H \tau-\frac{1}{2}\hat H^2 \tau^2+\mathcal O(\tau^3).
\end{equation}
Therefore, the wave function for short-time evolution is
\begin{equation}
\left| \psi(\tau) \right>=(I-i\hat H \tau-\frac{1}{2}\hat H^2 \tau^2)\left| \psi_{\mathrm{initial}} \right>+\mathcal O(\tau^3).
\end{equation}
Thus, the expansion coefficients of the wave function $\left| \psi(\tau) \right>=\sum_n c_n(\tau)\left| n \right>$ are:
\begin{equation}
    \begin{aligned}
        c_1(\tau) &= 4it_4 \tau \sin k_x \sin k_y + 2t_4 \tau^2 \sin k_x \sin k_y \left(3\varepsilon_{xx}(\mathbf{k}) + \varepsilon_{yy}(\mathbf{k}) + \delta U\right) + \mathcal{O}(\tau^3), \\
        c_2(\tau) &= 4it_4 \tau \sin k_x \sin k_y + 2t_4 \tau^2 \sin k_x \sin k_y \left(3\varepsilon_{xx}(\mathbf{k}) + \varepsilon_{yy}(\mathbf{k}) + \delta U\right) + \mathcal{O}(\tau^3), \\
        c_3(\tau) &= 1 - i\tau\left(2\varepsilon_{xx}(\mathbf{k}) + \delta U\right) - \tau^2\left[16t_4^2\sin^2 k_x\sin^2 k_y + 2\varepsilon_{xx}^2(\mathbf{k}) + 2\varepsilon_{xx}(\mathbf{k})\delta U + \frac{(\delta U)^2}{2}\right] + \mathcal{O}(\tau^3), \\
        c_4(\tau) &= -16t_4^2\sin^2 k_x\sin^2 k_y \cdot \tau^2 + \mathcal{O}(\tau^3).
    \end{aligned}
\end{equation}
It is not difficult to prove that the above expansion coefficients are automatically normalized to the order of $\tau^2$.
Using the expression for the orbital order, we obtain
\begin{equation}
O(\mathbf{k}, \tau) = 2 - 4\left(-4t_4 \sin k_x \sin k_y\right)^2 \tau^2 + \mathcal{O}(\tau^3) = 2 - 64t_4^2 \sin^2 k_x \sin^2 k_y \tau^2 + \mathcal{O}(\tau^3),
\end{equation}
and the global orbital order is
\begin{align}
O(\tau) = 2 - 16t_4^2 \tau^2.
\end{align}
Therefore, we find that in the short-time evolution process, only the hopping parameter $t_4$ between the two orbits affects the evolution rate, and the orbital order evolves according to the $\tau^2$ scaling.
To calculate the dynamics of entanglement entropy, we need to calculate $|c_n(\tau)|^2$ to the order of $\tau^2$:
\begin{equation}
    \begin{aligned}
    |c_1|^2 &= |c_2|^2 = 16t_4^2 \tau^2 \sin^2 k_x \sin^2 k_y + \mathcal{O}(\tau^3), \\
    |c_3|^2 &= 1 - 32t_4^2 \tau^2 \sin^2 k_x \sin^2 k_y + \mathcal{O}(\tau^3), \\
    |c_4|^2 &= \mathcal{O}(\tau^4).
\end{aligned}
\end{equation}
In the short-time evolution limit $\tau \rightarrow 0^+$,
\begin{align}
S(\mathbf{k}, \tau)= -64t_4^2 \tau^2 f(\mathbf{k}) \ln \tau - 32t_4^2 \tau^2 f(\mathbf{k}) \ln(16t_4^2 f(\mathbf{k})) + 16t_4^2 \tau^2 f(\mathbf{k}).
\end{align}
Here we define $f(\mathbf{k}) = \sin^2 k_x \sin^2 k_y$. The average entanglement entropy is
\begin{align}
S(\tau) &\approx -16t_4^2 \tau^2 \ln(2t_4 \tau) - 12t_4^2 \tau^2.
\end{align}
Therefore, in the very short time evolution, the entanglement entropy evolves with both $\tau^2$ and $-\tau^2 \ln \tau$ contributions. Initially, the $-\tau^2 \ln \tau$ term dominates, while the $\tau^2$ contribution becomes more significant as time evolves. Here, the evolution of entanglement entropy is also entirely determined by the hopping parameter between the two orbits.
\begin{figure}[H]  
    \centering
    \includegraphics[width=0.9\textwidth]{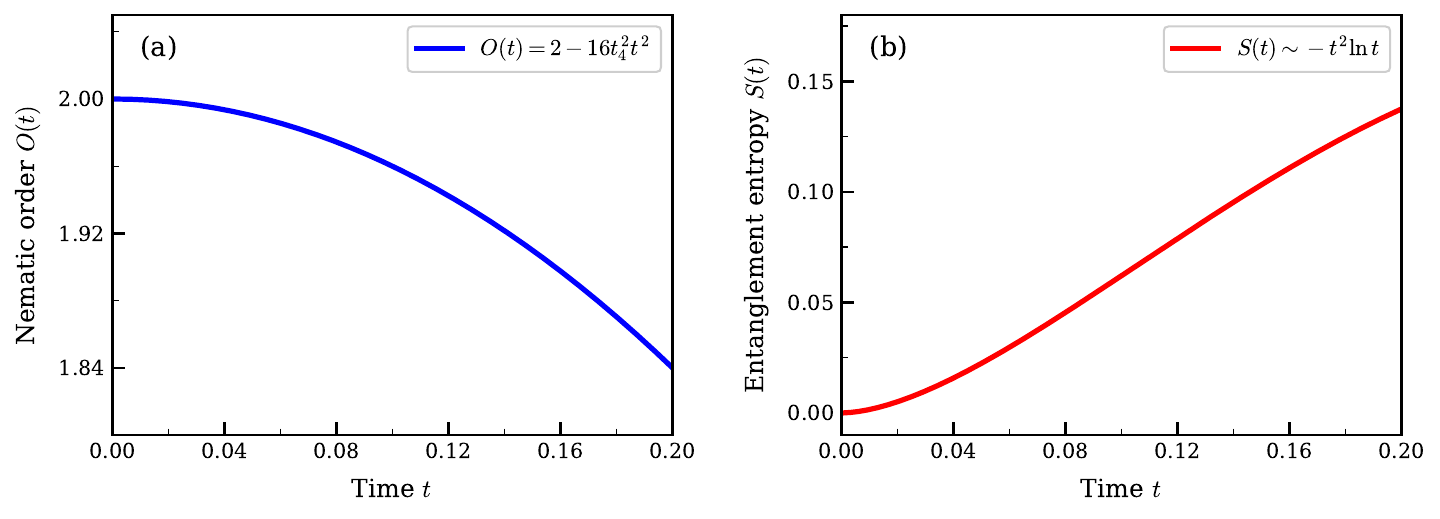}  
    \caption{Dynamics of physical quantities in the short-time regime. 
    (a) shows the evolution of the orbital order parameter $O(\tau)$ as a function of time $t$, 
    described by the formula $O(\tau) = 2 - 16t_4^2\tau^2$ with $t_4 = 0.5$; 
    (b) displays the time dependence of the entanglement entropy $S(\tau)$, follows the scaling $S(\tau) \sim -\tau^2 \ln \tau$.}
    \label{fig:quench_dynamics}  
\end{figure}

\section{Asymptotic Behavior of Entanglement Entropy and orbital Order}

In the long-time evolution process, both entanglement entropy and orbital order exhibit intense oscillations due to momentum-dependent phase factors, which usually cannot be solved analytically with precision. Therefore, we choose to use the method of stationary phase for asymptotic expansion. It is important to note that numerical calculations on finite lattices are insufficient to understand the evolution of long-time dynamics. This is because in a lattice system of size $L \times L$, the momentum space is discretized, and the minimum difference in momentum is $\Delta k_x = \frac{2\pi}{L}$. The dominant oscillating phase factor in the calculations of entanglement entropy or orbital order is typically determined by the energy gap $\Delta E(k_x, k_y)$: $e^{i\Delta E \tau}$. We expand the energy gap as follows:
\begin{equation}
\Delta E(k_x + \delta k_x, k_y + \delta k_y) = \Delta E(k_x, k_y) + \frac{\partial \Delta E}{\partial k_x}\delta k_x + \frac{\partial \Delta E}{\partial k_y}\delta k_y + \mathcal{O}(\delta k^2).
\end{equation}
Thus, the phase difference between nearby momenta $\delta \theta \sim \delta k_x \tau = \frac{2\pi \tau}{L}$. As $\tau$ increases, to accurately calculate the dynamics of physical quantities, a larger system size is required. Specifically, if we require $\delta \theta = \frac{2\pi}{M}$, where $M$ is an integer greater than 1, then the system size required at time $\tau$ is:
\begin{equation}
L \sim M \tau.
\end{equation}
However, in our problem, the computational complexity is $\mathcal{O}(L^2)$. In practical calculations, for a system size of $1200 \times 1200$, instability occurs at $\tau = 150$, which is completely insufficient for a rigorous asymptotic analysis of the scaling behavior during long-time evolution. Therefore, we must base our asymptotic analysis on analytical theory.

In the specific analysis, we utilize symmetry to categorize the parameters into four groups: (i) $t_1 = t_2, \delta U = 0$, (ii) $t_1 \neq t_2, \delta U = 0$, (iii) $t_1 = t_2, \delta U \neq 0$, and (iv) $t_1 \neq t_2, \delta U \neq 0$. The first three cases have analytical solutions for diagonalizing the Hamiltonian.

\subsection{$t_1 = t_2, \delta U = 0$}

\subsubsection{Asymptotic Behavior of Orbital Order Dynamics}

In this case, we denote $\varepsilon_{\mathrm{xx}}(\mathbf{k}) \equiv \varepsilon_d(\mathbf{k})$ and $\varepsilon_{\mathrm{xy}}(\mathbf{k}) \equiv \varepsilon_o(\mathbf{k})$. The Hamiltonian diagonalization results in the following eigenenergies:
\begin{equation}
E_1 = 2\varepsilon_d(\mathbf{k}) + 2\varepsilon_o(\mathbf{k}), \quad E_2 = 2\varepsilon_d(\mathbf{k}) - 2\varepsilon_o(\mathbf{k}), \quad E_3 = E_4 = 2\varepsilon_d(\mathbf{k}).
\end{equation}
The corresponding eigenstates are:
\begin{equation}
\psi_1 = \frac{1}{2}
\begin{bmatrix}
1 \\
1 \\
1 \\
1
\end{bmatrix}, \quad
\psi_2 = \frac{1}{2}
\begin{bmatrix}
1 \\
1 \\
-1 \\
-1
\end{bmatrix}, \quad
\psi_3 = \frac{1}{\sqrt{2}}
\begin{bmatrix}
1 \\
-1 \\
0 \\
0
\end{bmatrix}, \quad
\psi_4 = \frac{1}{\sqrt{2}}
\begin{bmatrix}
0 \\
0 \\
1 \\
-1
\end{bmatrix}.
\end{equation}
The momentum-dependent orbital order is:
\begin{align}
O(\mathbf{k}, \tau) &= 2\cos(2\varepsilon_o(\mathbf{k}) \tau).
\end{align}
Thus, the orbital order dynamics can be integrated as:
\begin{equation}
O(\tau) =  2 \left[ J_0(4 t_4 \tau) \right]^2.
\end{equation}
We find that the relaxation process can be characterized by the Bessel function. Using the properties of Bessel functions, we can analyze the short-time and long-time scaling behaviors of the relaxation process through asymptotic analysis.
When $\tau \rightarrow 0^+$, the series expansion of the zeroth-order Bessel function is:
\begin{equation}
J_0(z) = \sum_{m=0}^\infty \frac{(-1)^m}{(m!)^2} \left( \frac{z}{2} \right)^{2m} = 1 - \frac{z^2}{4} + \frac{z^4}{64} + \cdots.
\end{equation}
Thus, the leading order of the orbital order is:
\begin{equation}
O(\tau) \approx2 - 16 t_4^2 \tau^2.
\end{equation}
Therefore, the short-time scaling of the orbital order is $\tau^2$, which directly verifies our general calculation conclusion for the short-time process.
When the system evolves for a long time, $\tau \rightarrow \infty$, the asymptotic expansion of the zeroth-order Bessel function is:
\begin{equation}
J_0(z) \sim \sqrt{\frac{2}{\pi z}} \cos \left( z - \frac{\pi}{4} \right).
\end{equation}
\begin{figure}[H]  
    \centering
    \includegraphics[width=0.9\textwidth]{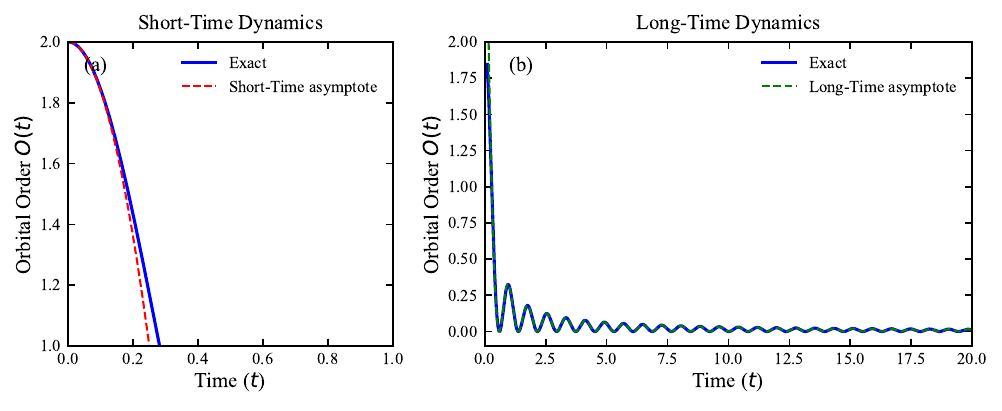}
    
    \caption{orbital order parameter dynamics. (a) shows the short-time behavior 
             comparing the exact solution (blue solid line) with the short-time asymptote 
             (red dashed line). (b) displays the long-time evolution with the exact solution 
             (blue solid line) and long-time asymptotic behavior (green dashed line). 
             Calculations use the hopping parameter $t_4 = 1.0$.}
    
    \label{fig:nematic_dynamics}
\end{figure}
Therefore, the long-time asymptotic expansion for the orbital order is:
\begin{equation}
    O(\tau)\approx\frac{1}{2\pi t_4\tau}(1+\sin(8t_4\tau)).
\end{equation}
\subsubsection{Asymptotic Behavior of Entanglement Entropy Dynamics}

We can calculate that at time $\tau$, the entanglement entropy has an analytical form:
\begin{equation}
S(\mathbf{k},\tau)=-4\bigl[\sin^{2}\phi\ln|\sin\phi|+\cos^{2}\phi\ln|\cos\phi|\bigr],
\end{equation}
where
\begin{equation}
\phi=4t_{4}\tau\,\lvert\sin k_{x}\sin k_{y}\rvert.
\end{equation}
\paragraph{Short-time limit ($\tau\to0^{+}$).}
When $\tau\to0^{+}$ $\phi\to0$, to leading order,
\begin{equation}
S(\mathbf{k},\tau)\approx-4\phi^{2}\ln\phi.
\end{equation}
Averaging over the Brillouin zone,
\begin{align}
\bar S(\tau)&\approx-16t_{4}^{2}\tau^{2}\Bigl[\ln(4t_{4}\tau)+\frac{1}{2}-\ln2\Bigr]
\sim16t_{4}^{2}\tau^{2}\ln\!\Bigl(\frac{1}{\tau}\Bigr),
\end{align}
confirming the general short-time prediction.

\paragraph{Long-time limit ($\tau\to\infty$).}
For fixed $\mathbf{k}$ with $\lvert\sin k_{x}\sin k_{y}\rvert>0$, $\phi\to\infty$, and $S(\mathbf{k},\tau)$ oscillates rapidly. Because $S(\phi)$ is periodic with period $\pi/2$,
\begin{equation}
S(\phi+\pi/2)=S(\phi),
\end{equation}
its time average equals the average over one period:
\begin{equation}
\bar S(\mathbf{k})=\frac{2}{\pi}\int_{0}^{\pi/2}\!S(\phi)\,d\phi=2(\ln4-1)\approx0.772.
\end{equation}
Since each momentum contributes independently, the steady-state average entanglement entropy is also $2(\ln4-1)$.

\paragraph{Approach to the steady state.}
We now discuss the dynamical evolution of entanglement entropy near its average value in the long-time limit $\tau \rightarrow \infty$. First, using the symmetry of the Brillouin zone, we restrict the integral range:
\begin{equation}
S(\tau) = \frac{1}{\pi^2} \iint_0^\pi S(\mathbf{k}, \tau) \, dk_x \, dk_y,
\end{equation}
where $S(\mathbf{k}, \tau) = -4[\sin^2 \phi \ln |\sin \phi| + \cos^2 \phi \ln |\cos \phi|]$, and $\phi = 4t_4 \tau |\sin k_x \sin k_y|$. For the long-time limit, slight changes in momentum in the Brillouin zone can cause significant changes in $\phi$, leading to violent oscillations in $S(k)$. The integral over momentum space will eventually lead to different momenta phases interfering to reach a steady-state value. Following the idea of the method of stationary phase, the main contribution to the integral comes from regions where the gradient of the phase factor is zero, i.e.,
\begin{equation}
    \begin{aligned}
    \nabla_{\mathbf{k}} \phi = 0 \Rightarrow
    \sin k_x \cos k_y &= 0, \\
    \cos k_x \sin k_y &= 0.
\end{aligned}
\end{equation}
We can solve for $(0,0), (\frac{\pi}{2}, \frac{\pi}{2}), (\pi, \pi), (0, \pi), (\pi, 0)$. Additionally, due to the absolute value, there are four curves that cannot be well-defined gradients, i.e., singular lines $k_x = 0, \pi, k_y = 0, \pi$. For these special regions, we can classify them into three categories: non-degenerate points, critical lines, and corner points.

\begin{figure}[H]  
    \centering
    \includegraphics[width=0.8\textwidth]{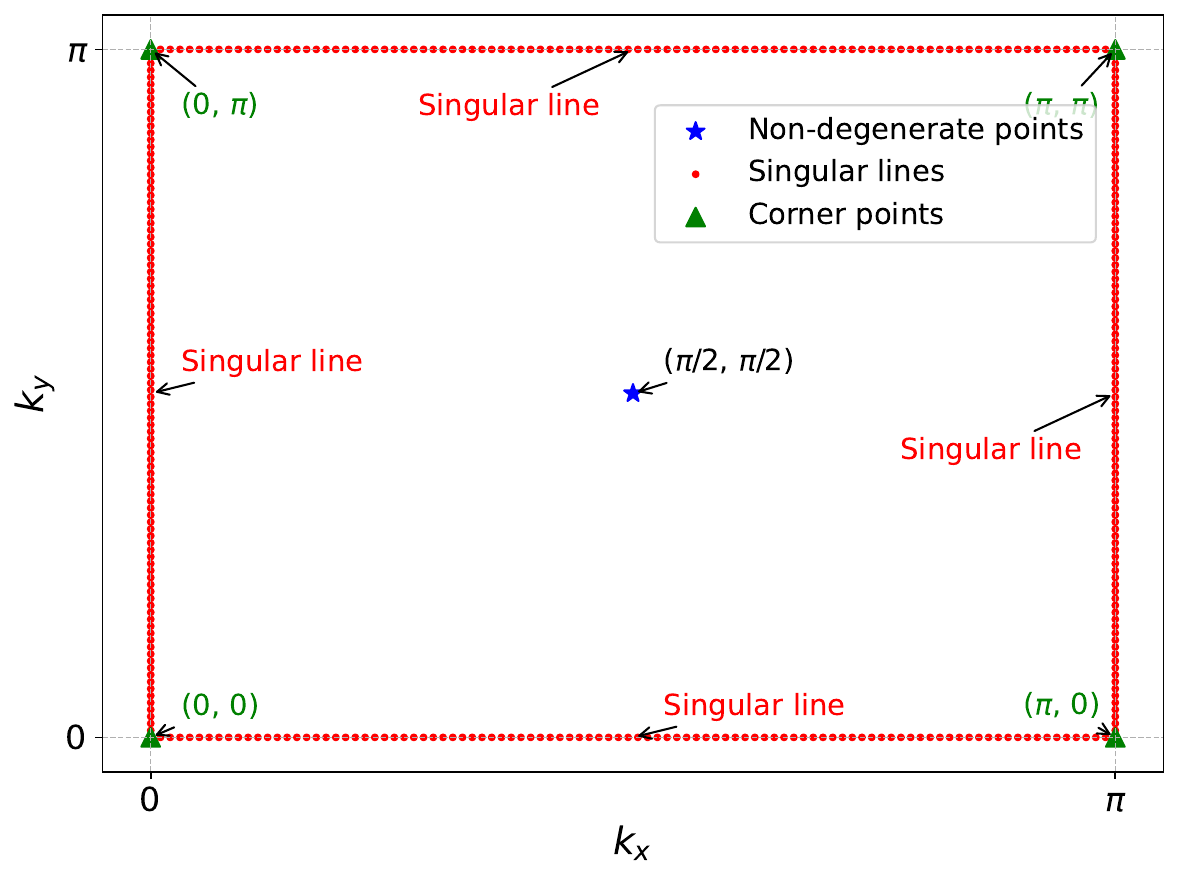}  
    
    \caption{Special points in the Brillouin zone. The blue star marks the non-degenerate point at $(\pi/2, \pi/2)$, 
             red dotted lines represent singular lines located at the boundaries $(k_x=0, k_x=\pi, k_y=0, k_y=\pi)$, 
             and green triangles indicate corner points at $(0,0)$, $(0,\pi)$, $(\pi,0)$, and $(\pi,\pi)$.}
    
    \label{fig:brillouin_zone}
\end{figure}

(i) Non-degenerate points: $(\frac{\pi}{2}, \frac{\pi}{2})$. At such points, the Hessian determinant
\begin{equation}
H_{ij} = \frac{\partial^2 \Delta E}{\partial k_i \partial k_j} \neq 0.
\end{equation}
The method of stationary phase indicates that near such points, an asymptotic expansion can be made:
\begin{equation}
\int e^{-i \Delta E_{ij}(\mathbf{k}) \tau} G_{ij}(\mathbf{k}) \, d^2 \mathbf{k} \sim \sum_{\text{stationary} \, \mathbf{k}^*} \frac{(2\pi)}{\tau \sqrt{|\det H|}} G_{ij}(\mathbf{k}^*) e^{-i \Delta E_{ij}(\mathbf{k}^*) \tau \pm i \pi/4}.
\end{equation}
Thus, it will provide an asymptotic contribution of $\frac{1}{\tau}$.

(ii) Corner points: $\mathbf{k} = (0,0), (0,\pi), (\pi,0), (\pi,\pi)$.
We take the point $(0,0)$ as an example and consider a nearby neighborhood $\delta$, let
\begin{equation}
u = k_x, \, v = k_y, \, (u, v \rightarrow 0).
\end{equation}
To ensure that the above approximation holds, we define the integral boundary region as $\phi < \epsilon$, then we have
\begin{equation}
uv < \frac{\epsilon}{4t_4 \tau}.
\end{equation}
Here $\epsilon \rightarrow 0^+$ is a small quantity. In this integral region, the integral function
\begin{equation}
S(\mathbf{k}, \tau) \approx -4 \phi^2 \ln \phi \approx -4 (4t_4 \tau uv)^2 \ln (4t_4 \tau uv).
\end{equation}
Now we calculate the integral
\begin{equation}
I_{\text{corner}} = \iint_{u>0, v>0, uv < \epsilon/(t_4 \tau)} -64 t_4^2 \tau^2 u^2 v^2 \ln (4 t_4 \tau uv) \, du \, dv.
\end{equation}
For convenience in calculation, define $c = 4t_4 \tau$, then the integral condition becomes $uv < \epsilon/c \equiv a$. The integral is rewritten as
\begin{equation}
I_{\text{corner}} = -4c^2 \iint_{u>0, v>0, uv < a} u^2 v^2 (\ln c + \ln u + \ln v) \, du \, dv.
\end{equation}
The result of this integral is
\begin{align}
I_{\text{corner}}&=-\frac{4\epsilon^{3}}{27}\ln\!\Bigl(\frac{\epsilon}{c\pi}\Bigr)
+\frac{8\epsilon^{3}}{81}
+\frac{4\epsilon^{3}}{3}\ln\!\Bigl(\frac{\epsilon}{c}\Bigr)\ln\!\Bigl(\frac{\pi^{2}c}{\epsilon}\Bigr)
-\frac{4\epsilon^{3}}{9}\ln\!\Bigl(\frac{\pi^{2}c}{\epsilon}\Bigr).
\end{align}
For fixed $\epsilon$ and asymptotically large $t$, the $\ln t/t$ term dominates:
\begin{equation}
I_{\text{corner}}\sim\frac{\epsilon^{3}}{9t_{4}\tau}\ln \tau+\mathcal O\!\Bigl(\frac{1}{\tau}\Bigr)
\;\sim\;\frac{\ln \tau}{\tau}.
\end{equation}

(iii) Critical lines $k_x=0,\pi,k_y=0,\pi$.
Specifically, we consider the critical line near $k_x=0$. On this critical line, $\phi=0$, hence the function $S(\mathbf{k},\tau)$ has a singularity. We consider the integral to exclude contributions from the corner points, i.e., $k_y\in[\delta,\pi-\delta]$.

To apply the long-time behavior, we perform a scaling on $k_x$,
\begin{equation}
k_x = \frac{u}{\sqrt{\tau}},
\end{equation}
and define $k_y = v$, with a constant coefficient $c = 4t_4$. The corresponding phase function is then
\begin{equation}
\phi \approx 4t_4 \sqrt{\tau} |u \sin v| = c \sqrt{\tau} |u \sin v|.
\end{equation}
Note that we are examining contributions from regions near the singular region $\phi \rightarrow 0$. Therefore, we need to constrain the integral region such that
\begin{equation}
\phi < \epsilon \Rightarrow c \sqrt{\tau} |u \sin v| < \epsilon \Rightarrow |u| < \frac{\epsilon}{c \sqrt{\tau} |\sin v|} \equiv b(v).
\end{equation}
Thus, the integral can be written as
\begin{equation}
I_{\text{line}} = \frac{1}{\pi^2} \tau^{-1/2} \int_{\delta}^{\pi-\delta} dv \int_0^{b(v)} f(c \sqrt{\tau} u \sin v) du.
\end{equation}
The contribution from the singular line is:
\begin{equation}
I_{\text{line}} \sim \frac{1}{\pi^2 c \tau} 2 \ln \left| \cot \frac{\delta}{2} \right| \left( -\frac{4}{3} \epsilon^3 \ln \epsilon + \frac{4}{9} \epsilon^3 \right) \sim \frac{1}{\tau}.
\end{equation}
Thus, we conclude that the contribution from the critical lines also decays as $1/\tau$.

\subsection{$t_1 \neq t_2, \delta U = 0$}

In this scenario, the Hamiltonian can be expressed as
\begin{equation}
h_{n_{\uparrow}=1,n_{\downarrow}=1}(k)=
\begin{bmatrix}
\varepsilon_{xx}(k)+\varepsilon_{yy}(k) & 0 & \varepsilon_{yx}(k) & \varepsilon_{xy}(k)\\
0 & \varepsilon_{xx}(k)+\varepsilon_{yy}(k) & \varepsilon_{yx}(k) & \varepsilon_{xy}(k)\\
\varepsilon_{xy}(k) & \varepsilon_{xy}(k) & 2\varepsilon_{xx}(k)& 0\\
\varepsilon_{yx}(k) & \varepsilon_{xy}(k) & 0 &2\varepsilon_{yy}(k)\\
\end{bmatrix}.
\end{equation}
We consider the non-symmetric parameters: $t_1=-1, t_2=1.3, t_3=t_4=-0.85$. The obtained eigenvalues are:
\begin{equation}
E_1=E_2=\varepsilon_{xx}(k)+\varepsilon_{yy}(k), E_{\pm}(k)=\varepsilon_{xx}(k)+\varepsilon_{yy}(k)\pm\Delta(k),
\end{equation}
where
\begin{equation}
\Delta(k)=\sqrt{\left( \varepsilon_{xx}(k)-\varepsilon_{yy}(k) \right)^2+4(\varepsilon_{xy}(k))^2}.
\end{equation}
Thus, the energy level structure consists of two doubly degenerate levels and two levels with a gap. At $\Delta(k)=0$, the energy level structure changes abruptly to fourfold degeneracy. 
\subsubsection{Asymptotic Behavior of Entanglement Entropy Dynamics}
The coefficients of the expansion are generally given by
\begin{equation}
c_n(\mathbf{k}, \tau) = \sum_i e^{-iE_i \tau} v_{i3} v_{in},
\end{equation}
and the entanglement entropy depends on $|c_n|^2$, thus it will include oscillating factors $e^{-i\Delta E_{ij}(\mathbf{k}) \tau}$, where $\Delta E_{ij} = E_i - E_j$ are the energy differences. In this problem, there exist three energy differences: $0, \Delta, 2\Delta$, hence generally, $|c_n|^2 = a_n(\mathbf{k}) + b_n(\mathbf{k}) \cos(\Delta(\mathbf{k}) \tau) + c_n(\mathbf{k}) \cos(2\Delta(\mathbf{k}) \tau)$, thus the entanglement entropy at a $\mathbf{k}$ point can be written as
\begin{equation}
    \begin{aligned}
    S(\mathbf{k}, \tau) =& -\sum_n \left(a_n(\mathbf{k}) + b_n(\mathbf{k}) \cos(\Delta(\mathbf{k}) \tau) + c_n(\mathbf{k}) \cos(2\Delta(\mathbf{k}) \tau)\right)\\
    &\ln\left(a_n(\mathbf{k})+ b_n(\mathbf{k}) \cos(\Delta(\mathbf{k}) \tau) + c_n(\mathbf{k}) \cos(2\Delta(\mathbf{k}) \tau)\right).
    \end{aligned}
\end{equation}
Since the entanglement entropy obviously has periodicity, we perform a Fourier series expansion:
\begin{equation}
S(\mathbf{k}, \tau) = \sum_{m=-\infty}^{\infty} s_m e^{im\Delta(\mathbf{k})\tau}
.
\end{equation}
Thus, the original integral becomes
\begin{equation}
S(\tau) = \sum_{m=-\infty}^{\infty} \iint_{[-\pi, \pi]^2} s_m(\mathbf{k}) e^{im\Delta \tau} dk_x dk_y \equiv \sum_{m=-\infty}^{\infty} S_m e^{im\Delta \tau},
\end{equation}
where we define
\begin{equation}
S_m = \iint_{[-\pi, \pi]^2} s_m(\mathbf{k}) dk_x dk_y.
\end{equation}
The $m=0$ component contributes to the non-oscillatory steady-state value, while the $m \neq 0$ parts contribute to the time-dependent oscillatory effects. It is evident that, in the long-time limit $\tau \rightarrow \infty$, the integral is primarily dominated by the stationary points, other regions cancel out due to rapid oscillations. The stationary points for the phase function $h_{ij}(\mathbf{k}) = \Delta E_{ij}(\mathbf{k})$ are determined by the condition:
\begin{equation}
\nabla_{\mathbf{k}}\Delta E_{ij} = 0.
\end{equation}
In this system, the energy differences are
\begin{align}
\Delta(\mathbf{k}) &= \sqrt{\left( \varepsilon_{xx}(\mathbf{k}) - \varepsilon_{yy}(\mathbf{k}) \right)^2 + 4 (\varepsilon_{xy}(\mathbf{k}))^2} \\
&= \sqrt{4[(t_1 - t_2)(\cos k_x - \cos k_y)]^2 + 16 t_4^2 \sin k_x^2 \sin k_y^2 }.
\end{align}
The type of the stationary points is determined by whether the Hessian determinant is zero:
\begin{equation}
H_{ij} = \frac{\partial^2 \Delta E}{\partial k_i \partial k_j}.
\end{equation}
Accordingly, we calculate and obtain five stationary points, the conditions are as follows:

\begin{center}
\begin{tabular}{|c|c|c|c|c|}
\hline
label & momentum & $\Delta E$ & Hessian determinant & Type \\
\hline
1 & $(0,0)$ & 0 & 0 & degenerate \\
2 & $(0,\pi)$ & $\neq 0$ & $\neq 0$ & non-degenerate \\
3 & $(\pi,0)$ & $\neq 0$ & $\neq 0$ & non-degenerate \\
4 & $(\pi,\pi)$ & 0 & 0 & degenerate \\
5 & $(\pi/2,\pi/2)$ & $\neq 0$ & $\neq 0$ & non-degenerate \\
\hline
\end{tabular}
\end{center}
For non-degenerate stationary points, there is an asymptotic expansion
\begin{equation}
\int e^{-i \Delta E_{ij}(\mathbf{k}) \tau} G_{ij}(\mathbf{k}) d^2 \mathbf{k} \sim \sum_{\text{stationary} \, \mathbf{k}^*} \frac{(2\pi)}{\tau \sqrt{|\det H|}} G_{ij}(\mathbf{k}^*) e^{-i \Delta E_{ij}(\mathbf{k}^*) \tau \pm i \pi/4}.
\end{equation}
Thus, the momenta $(0,\pi),(\pi,0),(\pi/2,\pi/2)$ will contribute to the $1/\tau$ scaling behavior, specifically
\begin{equation}
S_m(\tau) + S_{-m}(\tau) \sim \frac{4\pi}{|m| \tau} \sum_{j} \frac{\sin(m\tau \phi_j)}{\sqrt{|\det H_{\Delta,j}|}}.
\end{equation}
This means that the total entanglement entropy exhibits oscillations with multiple frequencies, but mainly dominated by those low frequencies (i.e., small $m$), and the amplitude decreases at a rate of $1/\tau$.

For degenerate stationary points, we need to perform a higher-order expansion. Taking $(0,0)$ as an example, near this point, we use polar coordinates to rewrite the gap as:
\begin{align}
\Delta(k_x,k_y)&=\sqrt{4\bigl[(t_1-t_2)(\cos k_x-\cos k_y)\bigr]^2+16t_4^2\sin^2k_x\sin^2k_y}\nonumber\\
&\approx 2\sqrt{\bigl[(t_1-t_2)(k_x^2-k_y^2)\bigr]^2+4t_4^2k_x^2k_y^2}\nonumber\\
&=2r^2\sqrt{(t_1-t_2)^2\cos^2 2\theta+t_4^2\sin^2 2\theta}\nonumber\\
&\equiv 2r^2A(\theta),\qquad
A(\theta):=\sqrt{(t_1-t_2)^2\cos^2 2\theta+t_4^2\sin^2 2\theta}.
\end{align}

Define the cut-off radius
\begin{equation}
\delta=\Bigl(\frac{\epsilon}{m\tau|B_{\max}|}\Bigr)^{1/4}\equiv C \tau^{-1/4},
\end{equation}
so that the quartic error term satisfies
$|\delta\phi|=m\tau|B_{\max}|\delta^4<\epsilon$.
The radial integral is
\begin{align}
\mathcal R(\tau,\theta)&=\int_0^{\delta}e^{i2m\tau A\tau (\theta)r^2}r\,dr\nonumber\\
&=\frac{1}{2}\int_0^{\delta^2}e^{i2m\tau A(\theta)\rho}\,d\rho\nonumber\\
&=\frac{1}{4im\tau A(\theta)}\Bigl(e^{i2m\tau A(\theta)\delta^2}-1\Bigr).
\end{align}

The degenerate-point contribution is therefore
\begin{align}
S_m^{\mathrm{deg}}(\tau)
&\approx s_m(0)\int_0^{2\pi}\!\!d\theta\,\mathcal R(\tau,\theta)\nonumber\\
&=\frac{s_m(0)}{4im\tau}\int_0^{2\pi}\frac{d\theta}{A(\theta)}
-\frac{s_m(0)}{4im\tau}\int_0^{2\pi}\frac{e^{i2m\tau A(\theta)\delta^2}}{A(\theta)}\,d\theta\nonumber\\
&\equiv S_{m;c}^{\mathrm{deg}}+S_{m;o}^{\mathrm{deg}}.
\end{align}

The constant (non-oscillatory) part decays as
\begin{equation}
S_{m;c}^{\mathrm{deg}}=-\frac{s_m(0)}{4im\tau}\int_0^{2\pi}\frac{d\theta}{A(\theta)}\sim\frac{1}{\tau},
\end{equation}
while the oscillatory part, by the Riemann–Lebesgue lemma, decays faster:
\begin{equation}
S_{m;o}^{\mathrm{deg}}=\mathcal O(\tau^{-5/4}),
\end{equation}
since the integral satisfies
\begin{equation}
\int_0^{2\pi}\frac{e^{i2mC^2A(\theta)\tau^{1/2}}}{A(\theta)}\,d\theta=\mathcal O(\tau^{-1/2}).
\end{equation}
Hence the total contribution from the degenerate points is
\begin{equation}
S^{\mathrm{deg}}(\tau)=\mathcal O(\tau^{-1})+\mathcal O(\tau^{-5/4})=\mathcal O(\tau^{-1}).
\end{equation}


Collecting the contributions from all stationary points (non-degenerate and degenerate), we obtain
\begin{equation}
S(\tau)=S_{\infty}+\frac{C}{\tau}+\mathcal O(\tau^{-5/4}),
\end{equation}
where $S_{\infty}$ is the steady-state value and $C$ is a constant independent of time.  The same analysis gives identical scaling for the orbital order:
\begin{equation}
O(\tau)=O_{\infty}+\frac{C'}{\tau}+\mathcal O(\tau^{-5/4}).
\end{equation}
\subsubsection{Dynamics of Orbital Order Asymptotic Behavior}

From the perspective of energy level structure, the orbital order can generally be expressed as
\begin{equation}
O(\mathbf{k},\tau) = A + B \cos \Delta(\mathbf{k}) \tau + C \cos 2 \Delta(\mathbf{k}) \tau,
\end{equation}
or written as:
\begin{equation}
O(\mathbf{k},\tau) = \sum_{m=-\infty}^{\infty} o_{m}(\mathbf{k},\tau) e^{im \Delta(\mathbf{k})\tau}.
\end{equation}
In the long-time evolution process $\tau \rightarrow \infty$, it is not difficult to find that at this time the orbital order will have the same asymptotic behavior as entanglement entropy, i.e.,
\begin{equation}
O(\tau) \sim \frac{1}{\tau}.
\end{equation}

\section{$t_1 = t_2, \delta U \neq 0$}

In this case, the Hamiltonian can be written as
\begin{equation}
h_{n_{\uparrow}=1,n_{\downarrow}=1}(k)=
\begin{bmatrix}
2\varepsilon_{xx}(k) & 0 & \varepsilon_{yx}(k) & \varepsilon_{xy}(k)\\
0 & 2\varepsilon_{xx}(k)& \varepsilon_{yx}(k) & \varepsilon_{xy}(k)\\
\varepsilon_{xy}(k) & \varepsilon_{xy}(k) & 2\varepsilon_{xx}(k)+\delta U & 0\\
\varepsilon_{yx}(k) & \varepsilon_{xy}(k) & 0 &2\varepsilon_{xx}(k)+\delta U\\
\end{bmatrix}.
\end{equation}
This Hamiltonian has non-degenerate eigenvalues
\begin{equation}
E_1(k) = 2\varepsilon_{xx}(k), \quad E_2(k) = 2\varepsilon_{xx}(k) + \delta U, \quad E_{\pm}(k) = 2\varepsilon(k) + \frac{\delta U}{2} \pm \frac{1}{2} \Delta(k),
\end{equation}
where
\begin{equation}
\Delta(k) = \sqrt{\delta U^2 + 16 \varepsilon_{xy}(k)^2}, \quad \varepsilon(k) = \varepsilon_{xx}(k).
\end{equation}

\subsubsection{Dynamics of Entanglement Entropy and Orbital Order Asymptotic Analysis}
In the long-time evolution process, the entanglement entropy's asymptotic behavior mainly depends on the structure of the energy gap. Here the energy gap has the form:
\begin{equation}
\Delta(k) = \sqrt{\delta U^2 + 16\varepsilon_{xy}(k)^2}.
\end{equation}
Using $\nabla_k\Delta=0$ to calculate the stationary point conditions:
\begin{equation}
    \begin{aligned}
    &\frac{64t_4\varepsilon_{xy}}{\Delta}\cos k_x \sin k_y=0,\\
    &\frac{64t_4\varepsilon_{xy}}{\Delta}\sin k_x \cos k_y=0.
\end{aligned}
\end{equation}
We can find two types of stationary points, the first type is one-dimensional line:
\begin{equation}
\gamma(\mathbf{k})=0 \Rightarrow k_x=0,\pi,\text{or},k_y=0,\pi.
\end{equation}
The second type is isolated stationary points:
\begin{equation}
(\pm\frac{\pi}{2},\pm\frac{\pi}{2}).
\end{equation}
It is straightforward to calculate that there are no degenerate points in this problem, so we can apply the method of stationary phase directly for the calculation. Among them, isolated stationary points will give an asymptotic behavior of $\tau$ but the situation for one-dimensional lines is different.
We consider the neighborhood of the curve stationary manifold, for example, the neighborhood of $(0,k_x)$, with a small amount $k_x=\delta k$ and $k_y$ fixed $(\sin k_y\neq0)$, leading to the approximation
\begin{equation}
\varepsilon_{xy}(\mathbf{k}) \approx -4t_4 \delta k \sin k_y.
\end{equation}
In the neighborhood of the curve stationary point, the integral becomes
\begin{align}
S_{\mathrm{boundary}}(\tau) &\approx \int_{-\epsilon}^{\epsilon} \int_{-\pi}^{\pi} f(0,k_y)e^{i[1+\alpha(k_y)\delta k^2]\tau}d\delta kdk_y\\
&\sim \frac{C}{\sqrt{\tau}}.
\end{align}
Therefore, the asymptotic behavior of the entanglement entropy is
\begin{equation}
S(\tau)\sim S_{\infty}+\frac{C}{\sqrt{\tau}}+\mathcal{O}(\frac{1}{\tau}).
\end{equation}
In general, the asymptotic behavior of the orbital order is also determined by the energy gap $\Delta(\mathbf{k})$, and through the same asymptotic analysis as above, we can obtain
\begin{equation}
O(\tau)\sim O_{\infty}+\frac{C}{\sqrt{\tau}}+\mathcal{O}(\frac{1}{\tau}).
\end{equation}

\section{$t_1 \neq t_2, \delta U \neq 0$}

In this case, the dynamics of orbital order and entanglement entropy generally depend on the energy level structure. Specifically, the energy gap $\Delta E(\mathbf{k})$ determines the phase dynamics during the evolution, and in the method of stationary phase, the asymptotic behavior of the integrals depends entirely on those stationary points where
\begin{equation}
\nabla_{\mathbf{k}} \Delta E(\mathbf{k}) = 0.
\end{equation}
If these momentum points satisfy $\det H \neq 0$, we refer to them as non-degenerate points. Isolated non-degenerate points will always provide a scaling behavior of $1/\tau$. If $\det H = 0$, we obtain degenerate points; generally, the decay rate is slower at degenerate points, but the specific asymptotic behavior depends on the higher order expansion of the energy gap with respect to momentum, i.e.,
\begin{equation}
\Delta E(\mathbf{k}) \approx \Delta E(\mathbf{k_0}) + \sum_{k=2}^\infty \frac{1}{k!} \left( \delta k_x \frac{\partial}{\partial k_x} + \delta k_y \frac{\partial}{\partial k_y} \right)^k \Delta E\bigr|_{\mathbf{k_0}}.
\end{equation}
For the corresponding oscillatory integrals:
\begin{equation}
I(\tau) = \frac{1}{(2\pi)^d} \iint d^{d}k \,f(\mathbf{k}) e^{i \Delta(\mathbf{k})\tau},
\end{equation}
if the leading term in the expansion of the phase near the degenerate point is $\Delta E(\mathbf{k}) = a + \delta k_x^b + \delta k_y^c$, then we can define a scaling transformation by letting
\begin{equation}
u = \tau^{1/b} \delta k_x, \quad v = \tau^{1/c} \delta k_y,
\end{equation}
the integral transforms to
\begin{equation}
I(\tau) \sim \tau^{-1/b-1/c} e^{ia\tau} \iint du dv e^{i(u^b+v^c)} f(\tau^{-1/b}u,\tau^{-1/c}v) \sim \tau^{-1/b-1/c}.
\end{equation}
At this point, both entanglement entropy and orbital order will exhibit an asymptotic behavior of $\tau^{-1/b-1/c}$. It can be seen that the higher the power dependence of the energy gap on momentum near the degenerate point, the slower the decay. Suppose $b=c=4$, then it will give a scaling of $\tau^{-1/2}$.

If the non-degenerate stationary points of the system form a continuous curve, then the asymptotic behavior will also change, resulting in a $1/\sqrt{\tau}$ asymptotic behavior. Additionally, we need to consider whether the overall $|c_n|^2$ has singularities, just like in the case of $t_1=t_2,U=V$. If there are, we need to perform normalization near the singular points to handle the oscillatory integral, which often results in slower decay behavior than $1/\tau$, such as the appearance of a $\ln \tau$ factor.

Therefore, we can conclude that for the general case, we first need to obtain the momentum distribution and corresponding stationary points and singular points of the energy gap through numerical calculations. Then, based on the type of stationary points and the asymptotic situation near singular points, determine whether there are more important terms than $1/\tau$ in the long-time evolution process.

Finally, we organize and obtain:
\begin{center}
\begin{tabular}{|c|c|c|c|c|}
\hline
parameter & $t_1=t_2,U=V$ & $t_1=t_2,U\neq V$ & $t_1\neq t_2,U=V$ & $t_1\neq t_2,U\neq V$ \\
\hline
orbital order short-time  & $\tau^2$ & $\tau^2$ & $\tau^2$ & $\tau^2$ \\
orbital order long-time  & $\frac{1}{\tau}$ & $\frac{1}{\sqrt{\tau}}$ & $\frac{1}{\tau}$ & $\frac{1}{\tau}$ or slower \\
entanglement entropy short-time  & $-\tau^2\ln \tau$ & $-\tau^2\ln \tau$ & $-\tau^2\ln \tau$ & $-\tau^2\ln \tau$ \\
entanglement entropy long-time  & $\frac{\ln \tau}{\tau}$ & $\frac{1}{\sqrt{\tau}}$ & $\frac{1}{\tau}$ & $\frac{1}{\tau}$ or slower \\
\hline
\end{tabular}
\end{center}

\end{document}